\documentclass[a4,11pt]{cip-submit-2019}  
\usepackage{color,soul}
\usepackage{afterpage}

\usepackage{wrapfig}
\usepackage{wasysym}
\usepackage{enumitem}
\usepackage{adjustbox}
\usepackage{ragged2e}
\usepackage[svgnames,table]{xcolor}
\usepackage{tikz}
\usepackage{longtable}
\usepackage{changepage}
\usepackage{setspace}
\usepackage{hhline}
\usepackage{multicol}
\usepackage{tabto}
\usepackage{float}
\usepackage{multirow}
\usepackage{makecell}
\usepackage{hyperref}
\usepackage{mathptmx}
\DeclareSymbolFont{epsilon}{OML}{cmm}{m}{it}
\DeclareMathSymbol{\epsilon}{\mathord}{epsilon}{"0F}
\def\DD{D\kern-.7em\raise0.25ex\hbox{\char '55}\kern.33em}
\usepackage{hyperref}
\hypersetup{
	colorlinks=false,
	pdfborder={0 0 0}
}
\usepackage{graphicx,subfigure,wrapfig,epstopdf}
\def\Mr{\uppercase}
\usepackage{graphics}
\usepackage{wasysym}
\usepackage{amsmath,amsxtra,amssymb,latexsym,amscd,color,cite,bm}
\usepackage[mathscr]{eucal}
\usepackage{array}
\usepackage{multirow}
\usepackage{cite} 

\def\oc{$^{16}$O+$^{12}$C\ }

\def\cc{$^{12}$C+$^{12}$C\ }

\def\AA{nucleus-nucleus\ }

\def\doi#1{\href{http://dx.doi.org/#1}{doi:#1}}
\def\titles#1{\title{\large\bf\noindent #1}}
\def\authors#1{\author{\begin{flushleft}{#1}\end{flushleft}}}
\def\authord#1#2{\indent\Mr{#1}$^{#2}$}
\def\addressed#1#2{\\[1mm]\textit{$\!\!\!^{#1}$\indent#2}}

\def\Email{$^{\dagger}$}
\def\email#1{\bigskip\href{mailto:#1}{\!\!\Email\textit{E-mail:}~{#1}}\\[3mm]}
\def\PublicationInformation#1#2#3#4{\\[4mm]\href{mailto:#1}{\!\!\Email\textit{E-mail:}~{#1}}\\[3mm]
	\textit{\indent Received #2}\\[1mm]
	\textit{Accepted for publication~#3}\\[1mm]
	\textit{Published~#4}}
\def\Keywords#1{\\[.2cm] \textnormal{Keywords:~{#1}}.} 
\def\AND{$\text{\Small AND }$}
\def\and{$\text{\tiny AND }$}

\def\Classification#1{\\[.2cm] \textnormal{Classification numbers:~{#1}.}}  

\newcommand{\sout}[1]{\unskip}

\def\ed{
	\bibliographystyle{cip-sty-2019}
	\bibliography{references-database-name}
\end{document}
}
\begin{document}
	\Year{2021}
	\Page{1}\Endpage{14}
	\titles{Elastic and inelastic alpha transfer in the \oc scattering}
	\authors{
	\authord{\authord{Nguyen Tri Toan Phuc}{1,2,3}, Nguyen Hoang Phuc}{3} \Email \AND \space 
	\authord{Dao Tien Khoa}{3}
	\newline
	\addressed{1}{Department of Nuclear Physics, Faculty of Physics and Engineering Physics, 
	 \\ University of Science, Ho Chi Minh City, Vietnam.}
	\addressed{2}{Vietnam National University, Ho Chi Minh City, Vietnam.}
	\addressed{3}{Institute for Nuclear Science and Technology, VINATOM \\ 
	 179 Hoang Quoc Viet, Cau Giay, Hanoi, Vietnam}	
	\space 
\PublicationInformation{nguyenhoangphuc.phy@gmail.com}{}{11 May 2021}{9 July 2021}
	}
	\maketitle
	\markboth{Elastic and inelastic alpha transfer in the \oc scattering}{N.~T.~T. PHUC, 
	 N.~H. PHUC, AND \space D.~T. KHOA}

\begin{abstract}
The elastic scattering cross section measured at energies $E\lesssim 10$ MeV/nucleon 
for some light heavy-ion systems having two identical cores like \oc exhibits 
an enhanced oscillatory pattern at the backward angles. 
Such a pattern is known to be due to the transfer of the valence nucleon or cluster 
between the two identical cores. In particular, the elastic $\alpha$ transfer has been 
shown to originate directly from the core-exchange symmetry in the elastic \oc scattering. 
Given the strong transition strength of the $2^+_1$ state of $^{12}$C and its large 
overlap with the $^{16}$O ground state, it is natural to expect a similar $\alpha$ 
transfer process (or inelastic $\alpha$ transfer) to take place in the inelastic 
\oc scattering. The present work provides a realistic coupled channel description 
of the $\alpha$ transfer in the inelastic \oc scattering at low energies. Based on 
the results of the 4 coupled reaction-channels calculation, we show a significant 
contribution of the $\alpha$ transfer to the inelastic \oc scattering cross section 
at the backward angles. These results suggest that the explicit coupling to the 
$\alpha$ transfer channels is crucial in the studies of the elastic and inelastic 
scattering of a nucleus-nucleus system with the core-exchange symmetry.
\Classification{25.70.Hi, 21.60.Cs, 24.10.Eq}
\end{abstract} 

\section{\Mr{Introduction}}
For some light nucleus-nucleus systems having two identical cores, the measured 
elastic scattering cross section shows an enhanced oscillation at backward angles. 
Such a pattern is established to be due to the elastic transfer of the valence 
nucleon or cluster between the two identical cores, and the observed oscillations 
originate from the interference between the elastic scattering and transfer 
amplitudes. The elastic transfer process was observed for several light heavy-ion 
(HI) systems at low energies \cite{Brau82}, and it had been studied 
over the years to extract the structure information on the nucleon- and cluster 
spectroscopic factors (see, e.g., the review \cite{vOe75}) and the asymptotic 
normalization coefficients \cite{Mud97,Tri14}. Our recent studies have shown clearly 
the direct link between the elastic transfer process and the parity dependence 
of the nucleus-nucleus optical potential (OP), a natural consequence of the 
core-exchange symmetry of the dinuclear system under study \cite{Phuc19,Phuc21}. 
These results are an important step toward a deeper understanding 
of the core-exchange effect in the elastic nucleus-nucleus scattering, which has 
motivated a renewed interest on this topic \cite{Doh21}.   

The core-exchange symmetry and the associated transfer processes show up not only 
in the elastic scattering but also in the inelastic light HI scattering \cite{vOe75}, 
which is dubbed as the inelastic transfer process. Although several experimental 
\cite{Pot98,Sro83,Jar91,Jar92,Rud01,Chua78,Boh85,Voi88,Gel74,Kal75,vOe96}
and theoretical studies \cite{Ima87,Ima97,Bau74,Spa00} were done to investigate 
the inelastic nucleon transfer, the inelastic $\alpha$ transfer process was rarely 
studied so far. The finite-range distorted wave Born approximation (DWBA) and, 
more recently, coupled reaction channels (CRC) method have been used to explore 
the impact of the inelastic $\alpha$ transfer on the $^{16}$O+$^{12}$C 
\cite{DeV73,Boh78,Fer19}, $^{9}$Be+$^{13}$C \cite{Bar90}, $^{16}$O+$^{20}$Ne, and 
$^{14}$N+$^{10}$B scattering at low energies \cite{Mot79}. These studies show 
a significant contribution of the $\alpha$ transfer process to the inelastic scattering 
cross section at the backward angles, a situation similar to that of the elastic $\alpha$ 
transfer. At variance with the elastic transfer, the inelastic $\alpha$ transfer carries 
more structure information on the core excitation, in particular, the $\alpha$-core 
configuration of the excited state. A consistent CRC description of both the elastic 
and inelastic scattering with the explicit coupling to the elastic and inelastic $\alpha$ 
transfer channels must be a severe test of the models of the \AA OP, 
inelastic scattering- and transfer form factor (FF). Moreover, the $\alpha$ transfer 
channel seems to dominate over other transfer channels in some core-identical systems 
\cite{Phuc18}, so that to a good approximation the most important reaction channels 
belong to the same mass partition, and the number of unconstrained parameters can be 
reduced in the CRC calculation.            

The $\alpha$ transfer in the elastic \oc scattering at low energies is among the 
earliest elastic transfer reactions discovered \cite{vOe68}, and this is also the 
most studied case of the elastic $\alpha$ transfer. The elastic \oc scattering at 
energies $E\gtrsim 10$ MeV/nucleon is proven to be strongly refractive, with a pronounced 
nuclear rainbow pattern associated with a broad oscillation of the Airy minima 
observed in the elastic scattering cross section \cite{Kho16}. The observation
of the nuclear rainbow also enabled an accurate determination of the real \AA OP  
down to small internuclear distances \cite{Bra97,Kho07r}. In a recent study \cite{Phuc18}, 
we have carried out a systematic 10-channel CRC analysis of the elastic \oc scattering 
at the refractive energies and shown that the enhanced oscillation of the elastic cross 
section at the backward angles is due to the multistep $\alpha$ transfer processes. 
In particular, the indirect $\alpha$ transfer through the $2^+_1$ (4.44 MeV) excitation 
of $^{12}$C was shown to be the dominant transfer process in the elastic \oc scattering. 
Given a strong $E2$ coupling from the ground state to the $2^+_1$ state of $^{12}$C 
\cite{Kho08}, the inelastic $\alpha$ transfer is expected to contribute significantly 
to the inelastic \oc scattering to the $2^+_1$ state of $^{12}$C at the backward angles. 
We note here a recent CRC analysis using the algebraic cluster model \cite{Fer19} that 
found some effects of the inelastic $\alpha$ transfer in the inelastic \oc scattering 
data at $E_{\rm lab}=80$ \cite{Gut73,Boh78,Szi06} and 84 MeV \cite{Boh78}. However, 
the inelastic couplings to the low-lying excited states of $^{12}$C were found also 
strong at the backward angles, and the inelastic $\alpha$ transfer could not be 
determined unambiguously.       

To further explore the $\alpha$ transfer contribution to the inelastic \oc scattering, 
we perform in the present work a consistent CRC analysis of the elastic and inelastic 
\oc scattering data measured at the energies $E_\text{lab}=100, 115.9$, and 124 MeV 
at the Strasbourg Tandem Vivitron \cite{Nic00,Szi06,Szilner}. 
These inelastic \oc scattering data (for the $2^+_1$ state of $^{12}$C) were measured 
accurately up to the angles $\theta_\text{c.m.}$ beyond $100^\circ$. At such large 
angles, the $\alpha$ transfer amplitude is well separated from the pure inelastic 
scattering amplitude \cite{Phuc18} and can be unambiguously determined. 
Given the indirect $\alpha$ transfer via the $2^+_1$ excitation of the $^{12}$C 
core is the dominant process in the \oc scattering \cite{Phuc18,Fer19}, we restrict 
our CRC coupling scheme to 4 reaction channels, with the $^{12}$C core exchange 
in both the ground- and $2^+_1$ states explicitly taken into account. 
 
\section{\Mr{CRC formalism}}
\label{sec2}
For the core-identical \oc system, the elastic and inelastic $\alpha$ transfer channels 
are indistinguishable from the elastic and inelastic scattering channels, respectively. 
Therefore, the angular distributions of the observed elastic and inelastic \oc scattering 
are the coherent sums of the pure scattering and $\alpha$ transfer amplitudes, and
the total cross section is given by \cite{vOe75,Phuc19}
\begin{equation}
	\frac{d\sigma(\theta)}{d\Omega}=\left|f(\theta)\right|^2=\left|f_{\rm scat}(\theta)
	+f_{\rm trans}(\pi-\theta)\right|^2, \label{eq:fsig}
\end{equation}
where the $f_{\rm scat}$ and $f_{\rm trans}$ are the pure scattering and $\alpha$ 
transfer amplitudes of the elastic or inelastic scattering. Historically, the elastic 
and inelastic scattering amplitudes were often evaluated separately first, and then added 
to the corresponding transfer amplitudes obtained in the DWBA (see, e.g., 
Refs.~\cite{Bau74,Boh78}). With the introducing of the CRC model \cite{Sat83,Tho09}, 
the simultaneous description of the involved scattering and transfer channels can be 
achieved in a consistent and unified manner. 

\begin{figure}\vspace*{0cm}\hspace*{0cm}
	\includegraphics[width=0.65\textwidth]{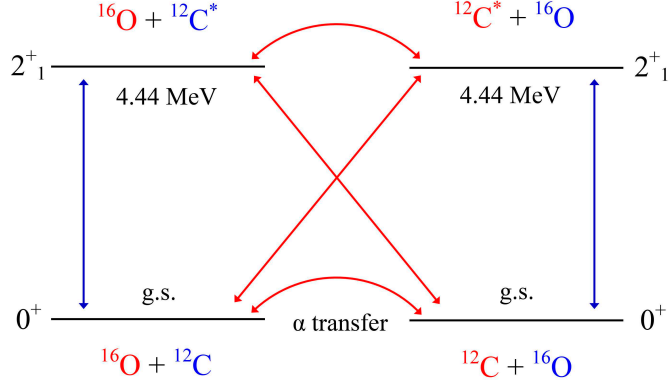}\vspace*{0cm}
	\caption{4-channel coupling scheme of the CRC calculation of the 
	elastic \oc scattering that includes both the direct and indirect (via the 
	2$^+_1$ excitation of the $^{12}$C core) $\alpha$ transfers.} \label{fig:Coup}
\end{figure}

We have performed recently a systematic CRC analysis of the elastic \oc scattering 
at $E_\text{lab}=100-300$ MeV \cite{Phuc18} taking into account up to 10 reaction
channels between the ground state and excited states of the $^{12}$C and $^{16}$O 
nuclei, with the $\alpha$ transfer treated explicitly. In addition to the direct 
elastic $\alpha$ transfer between $^{12}$C$_{\rm g.s.}$ and $^{16}$O$_{\rm g.s.}$,
we found a very strong contribution of the indirect $\alpha$ transfer via the $2^+_1$ 
excitation of $^{12}$C to the total elastic $\alpha$ transfer, and these two 
processes account mostly for the elastic $\alpha$ transfer cross section at the 
backward angles. This conclusion was confirmed by a later CRC study of the \oc 
scattering at lower energies \cite{Fer19}. Given the dominance of the elastic 
scattering and inelastic scattering to the $2^+_1$ state of $^{12}$C, we restrict 
the present CRC calculation of the \oc scattering to the 4-channel coupling 
scheme shown in Fig.~\ref{fig:Coup}, which is the smallest model space needed 
to describe simultaneously the elastic and inelastic \oc scattering, with the 
$\alpha$ transfer explicitly taken into account.        

In the CRC formalism, the coupled channel equation in the post form for a specific 
channel $\beta$ is given by \cite{Sat83,Tho09}
\begin{equation}
	(E_\beta-T_\beta-U_\beta)\chi_\beta=\sum_{\beta'\neq\beta,x=x'}
	\langle\beta|V|\beta'\rangle\chi_{\beta'}+\sum_{\beta'\neq\beta,x\neq x'}
	[\langle\beta|W_{\beta'}|\beta'\rangle +\langle\beta|\beta'\rangle
	(T_{\beta'}+U_{\beta'}-E_{\beta'})]\chi_{\beta'}, \label{eq:crc1}
\end{equation}
where $\beta'$ indicates a channel different from $\beta$, $x$ and $x'$ are the mass 
partitions associated with the scattering and $\alpha$-transfer channels. $E_\text{i}$ and $T_\text{i}$ are the asymptotic kinetic energy and kinetic energy operator of i-th channel in the centre-of-mass frame, respectively. $U_\beta$ and $U_{\beta'}$ are the (diagonal) OP, and $\chi_\beta$ and $\chi_{\beta'}$ are the corresponding relative-motion wave functions of the dinuclear system. $V$ is the local interaction operator and its matrix element $\langle\beta|V|\beta'\rangle$ is referred to as the form factor (FF). Due to the indistinguishability of the entrance and exit channels, the post- and prior forms of the coupled channel equation are the same, and the transition potential $W_\beta$ is determined \cite{Sat83,Tho09} by 
\begin{equation}
	W_\beta= V_{\alpha+^{12}\text{C}}+(U_{^{12}\text{C}+^{12}\text{C}}-
	U_{^{16}\text{O}+^{12}\text{C}}), \label{eq:crc2}
\end{equation}
with the complex remnant term $(U_{^{12}\text{C}+^{12}\text{C}}-U_{^{16}\text{O}+^{12}\text{C}})$ 
determined by the difference between the OP of the exit channel and the that between 
the two cores. $V_{\alpha+^{12}\text{C}}$ is the potential binding the $\alpha$ cluster 
to the $^{12}$C core in $^{16}$O. The CRC equations (\ref{eq:crc1}) for the 4 coupled 
reaction channels are solved iteratively using the code FRESCO written by Thompson \cite{Tho88}, 
with the non-orthogonality corrections and finite-range complex remnant terms properly 
taken into account. All the OP's have their real part calculated in the double-folding
model (DFM) \cite{Kho16} and the imaginary part parametrized in the Woods-Saxon (WS) form, 
which has the following form at the internuclear distance $R$
\begin{equation}
	U(R)=N_RU_{\rm F}(E,R)-\frac{iW_V }{1+\exp[(R-R_V)/a_V]}+V_{\rm C}(R).
	\label{eq:OP} 
\end{equation}  
The Coulomb potential $V_{\rm C}(R)$ is given by that of a point charge interacting 
with a uniform charge sphere \cite{Tho09} of the radius 
$R_{\rm ch}=0.95 \times (A_{\rm T}^{1/3}+A_{\rm P}^{1/3})$ (fm), 
where $A_{\rm T}$ and $A_{\rm P}$ are the target and projectile mass number, respectively. 
This choice of the Coulomb potential gives practically about the same result of the optical
model calculation as that using the more sophisticated one in Ref.~\cite{Pol76} for the 
elastic light HI scattering \cite{Sat83}. The real folded potential $U_{\rm F}(E,R)$ was 
calculated in the DFM using the CDM3Y3 density dependent NN interaction \cite{Kho97},
with the rearrangement term properly taken into account \cite{Kho16}. The ground state 
densities of $^{12}$C and $^{16}$O used in the DFM calculation were taken as the Fermi 
distributions with parameters adjusted to reproduce the empirical matter radii 
of these nuclei \cite{Kho01}. The renormalization factor $N_R$ of the real folded potential 
and WS parameters of the imaginary potential were adjusted to the best description 
of both the elastic and inelastic data at the forward angles up to 
$\theta_\text{c.m.}\approx 90^\circ$. The Airy minima observed in the elastic data 
were served as the important constraint for these parameters.    

The (complex) inelastic scattering FF for the 
$^{12}$C$_{\rm g.s.}\to ^{12}$C$_{2^+_1}$ transition is also obtained in the DFM \cite{Kho00} 
using the same CDM3Y3 interaction \cite{Kho16} with a complex density dependence 
suggested in Ref.~\cite{Kho08}, and the nuclear transition densities given by the 
$3\alpha$ resonating group method (RGM) \cite{Kam81}. The real part of the diagonal OP 
for the $U_{^{16}\text{O}+^{12}\text{C}(2^+_1)}$ system is obtained in the DFM calculation 
using the diagonal density of the 2$^+_1$ state taken from the RGM calculation, and 
the imaginary part is assumed to be the same as the ground state absorption. 

The $\alpha$-cluster structure of $^{16}$O enters the CRC calculation via the overlap 
function \cite{Sat83}, which is also known as the reduced-width amplitude \cite{Hor12}
\begin{equation}
	\langle{\rm ^{12}C}|^{16}{\rm O}\rangle=A_{NL}(^{16}{\rm O},{\rm ^{12}C})
	\Phi_{NL}(\bm{r}_{\alpha+{\rm ^{12}C}}), \label{eq:Amp0}
\end{equation}  
where $A_{NL}$ is the spectroscopic amplitude and $\Phi_{NL}(\bm{r}_{\alpha+{\rm ^{12}C}})$ 
is the relative-motion wave function of the $\alpha$+$^{12}$C system. This wave function 
is generated by the  binding potential $V_{\alpha+^{12}\text{C}}$ chosen in the WS form. 
In the present work, we adopt the WS geometry with $R=3.72$ fm and $a=0.845$ fm, 
parameterized in Ref.~\cite{Fukui19} based on the results of the five-body 
($^{12}$C plus 4 nucleons) calculation \cite{Hor14}. This five-body model is an extended 
version of the orthogonality condition model (OCM)  of $^{16}$O \cite{Suz76}, which
can describe accurately both the $0^+_1$ ground state and $0^+_2$ excited state 
of $^{16}$O. The WS potential based on the five-body model \cite{Hor14} has been used 
to reproduce with good accuracy the $\alpha$ transfer reaction 
$^{12}$C($^{6}$Li,$d$)$^{16}$O data at forward angles \cite{Fukui19}. 

Using the fixed WS geometry \cite{Fukui19}, the depth of the binding potential
$V_{\alpha+^{12}\text{C}}$ for $\Phi_{NL}(\bm{r}_{\alpha+{\rm ^{12}C}})$ is adjusted 
to reproduce the $\alpha$ separation energies of $^{16}$O, with the $^{12}$C core 
being in both the $0^+_1$ ground state and $2^+_1$ excited state
\begin{equation}
	E_\alpha = E_\alpha({\rm g.s.}) + E(^{12}{\rm C}^*),
	\label{eq:energy}
\end{equation}
where the $\alpha$ separation energy of $^{16}$O in the ground state is 
$E_\alpha({\rm g.s.})\approx 7.162$ MeV \cite{Til93}, and $E(^{12}{\rm C}^*)\approx 4.44$ MeV
is the excitation energy $E_{2^+_1}$ of $^{12}$C. In Eq.~(\ref{eq:Amp0}), the relative-motion 
wave function $\Phi_{NL}(\bm{r}_{\alpha+{\rm ^{12}C}})$ is characterized by the number 
of radial nodes $N$ and cluster orbital angular momentum $L$ that obey the Wildermuth's 
condition \cite{Sat83,Tho09}
\begin{equation}
	2(N-1)+L=\sum_{i=1}^{4}2(n_i-1)+l_i, \label{eq:Wildermuth}
\end{equation} 
where $l_i$ and $n_i$ are, respectively, the orbital angular momentum and principal 
quantum number of each constituent nucleon in the $\alpha$ cluster. Here we use the
number-of-nodes convention where the node at origin is included and the one at 
infinity is excluded. 

Similar to our previous CRC analysis \cite{Phuc18}, the $\alpha$ spectroscopic factors 
$S_\alpha=|A_{NL}|^2$ are taken from the results of the cluster-nucleon configuration 
interacting model \cite{Vol15,Vol17}. This large-scale shell model calculation is carried 
out in the unrestricted $psd$ model space and adopts an improved definition of $S_\alpha$
by Fliessbach \cite{Fli76,Fli77}. Although such a definition of $S_\alpha$ (also known as 
the amount of clustering) was used in microscopic cluster decay studies in the late 
nineties \cite{Lov98}, it has been used in the CRC study of the direct $\alpha$ transfer 
reaction only recently \cite{Phuc18}. A good agreement between the CRC results obtained
in Ref.~\cite{Phuc18} with the experimental data can serve as the validation for the use 
of this new $S_\alpha$ definition in other studies of the direct nuclear reactions. 

The use of the WS shape of the $\alpha$ binding potential and spectroscopic factors 
based on the reliable structure models reduces the uncertainty associated with 
the $\alpha$-cluster structure of $^{16}$O. For the ground state of $^{16}$O, we have 
$S_\alpha\approx 0.794$, $N=3$, and $L=0$ for the $\alpha + {\rm ^{12}C_{g.s.}}$ 
configuration, and $S^*_\alpha\approx 3.9$, $N=2$, and $L=2$ for  
$\alpha + {\rm ^{12}C_{2^+_1}}$. The $S^*_\alpha$ adopted for the 
$\alpha + {\rm ^{12}C_{2^+_1}}$ configuration is nearly 5 times large than 
that adopted for the $\alpha + {\rm ^{12}C_{g.s.}}$ configuration, which is due 
mainly to the $M$-substate degeneracy \cite{Ich73}. The ratio 
$S^*_\alpha/S_\alpha\approx 5$ is consistent with the results of different 
structure models as discussed in Ref.~\cite{Phuc18}. We note that the value $S^*_\alpha\approx 3.9$ calculated by the shell model \cite{Vol15,Vol17} for core-excited configuration is larger than those reported in the literature (see, e.g, Ref.~\cite{Phuc18}). This is a direct consequence of the Fliessbach's definition of $\alpha$ spectroscopic factors \cite{Fli76,Fli77} used in Ref.~\cite{Vol15,Vol17}, which properly takes into account the orthonormalization and antisymmetrization for the clustering channel.  

\section{\Mr{Elastic and inelastic \oc scattering with alpha transfer}}
\label{sec3}
\begin{figure}[bhtp!]\vspace*{-1cm}\hspace*{0cm}
	\includegraphics[width=0.8\textwidth]{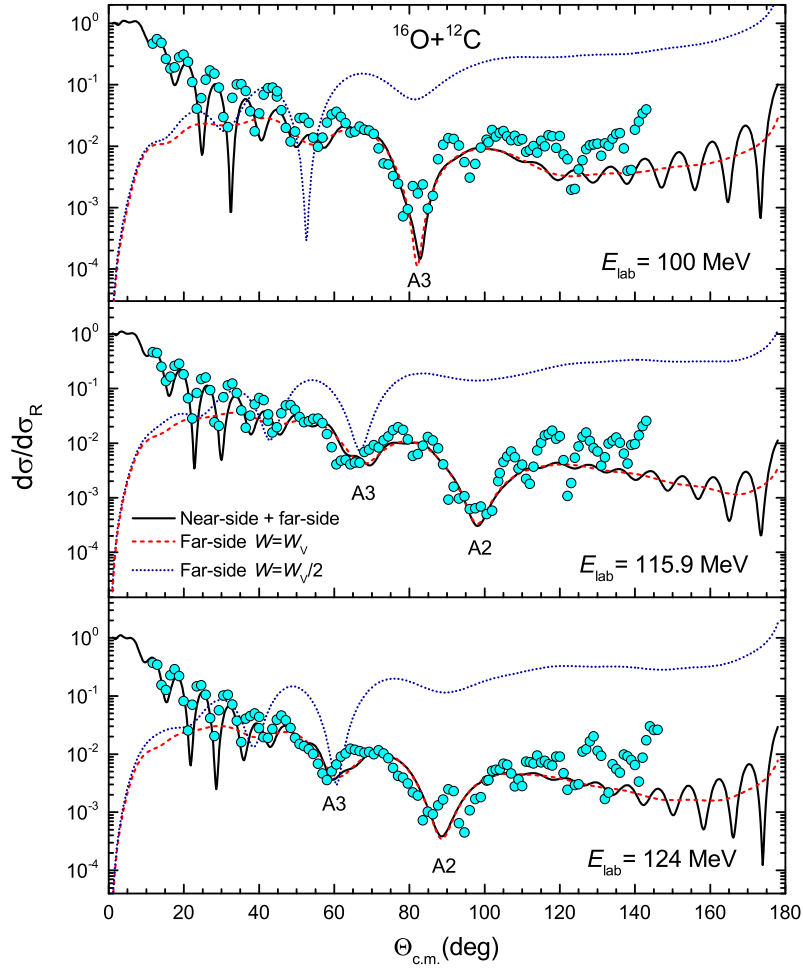}\vspace*{-1.5cm}
	\caption{CC description (solid lines) of the purely elastic \oc scattering 
  in comparison with the data	measured at $E_\text{lab}=100, 115.9$, and 124 MeV 
	\cite{Nic00}. The cross section of the far-side scattering has been obtained 
	with 2 different strengths of the absorption (see parameters of the WS 
	imaginary OP in Table \ref{t:OP}) using the Fuller's method \cite{Ful75}.} 
 \label{fig:ES}
\end{figure}

Given the significant $\alpha$ spectroscopic factors of the $\alpha + {\rm ^{12}C}$ 
configurations of the $^{16}$O nucleus discussed above and strong coupling effect 
of the $\alpha$-cluster states shown by the structure calculations 
\cite{Epe14,Bij14,Enyo17,Wang19}, some contributions of the $\alpha$ transfer 
channels to the elastic \oc cross section are naturally expected at backward angles. 
A realistic description of the purely elastic scattering is very important for 
the present CRC study of the $\alpha$ transfer in the \oc scattering. We consider 
here the elastic \oc scattering data measured accurately at the energies $E_\text{lab}=
100, 115.9$, and 124 MeV by the Strasbourg group \cite{Nic00}, approaching the range 
of refractive energies for the \oc system \cite{Kho16}. 
\begin{table}[bht]
	\caption{Parameters (\ref{eq:OP}) of the complex OP used in the 4-channel CRC 
	calculation of the elastic and inelastic \oc scatterings at $E_\text{lab}=100, 115.9$, 
	and 124 MeV. $N_R$ is the renormalization factor of the real folded OP, $J_R$ and 
	$J_W$ are the volume integrals of the real and imaginary parts of the OP, 
	respectively. $\sigma_R$ is the total reaction cross section.} 	\label{t:OP} 
	\begin{tabular}{|c|c|c|c|c|c|c|c|} \hline
		$E_{\rm lab}$ & $N_R$ & $J_R$ & $W_V$ & $R_V$ & $a_V$ & $J_W$ & $\sigma_R$ \\
		(MeV) &  & (MeV~fm$^3$) & (MeV) & (fm) & (fm) & (MeV~fm$^3$) & (mb) \\ \hline
		100   & 0.943 & 311.3 &	11.48 &	5.67 &	0.47  &	48.6  &	1305  \\
		115.9 & 0.950 & 311.7 &	12.80 & 5.70 &	0.46  &	55.0  &	1328  \\ 		
		124   & 0.950 & 310.8 &	12.84 &	5.64 &	0.58  &	55.4  &	1410  \\ \hline  
	\end{tabular}
\end{table}
The refractive nature of the elastic \oc scattering \cite{Kho16} is very well illustrated 
by the near-far decomposition of the elastic scattering amplitude based on the method 
developed by Fuller \cite{Ful75}. Namely, by decomposing the scattering amplitude into those 
representing the two waves traveling in $\theta$ that are running in the opposite 
directions around the scattering center, the elastic amplitude $f(\theta)$ can be 
determined as a sum of the near-side ($f_{\rm N}$) and far-side ($f_{\rm F}$) amplitudes
\begin{equation}
  f(\theta)=f_{\rm N}(\theta)+f_{\rm F}(\theta).
 \label{NFdec}
 \end{equation} 
The explicit expressions of these two amplitudes are discussed, e.g., in 
Refs.~\cite{Kho07r,Phuc18}. We emphasize here that $f_{\rm N}(\theta)$ represents 
the waves deflected to the direction of $\theta$ on the near side of the scattering 
center, and the waves traveling on the opposite, far side of the scattering center 
to the same angle $\theta$ give rise to $f_{\rm F}(\theta)$. Therefore, the \emph{diffractive}
near-side scattering occurs mainly at the surface of the two colliding nuclei, while the  
\emph{refractive} far-side scattering penetrates more into the interior of the \AA system. 
The broad oscillation of the far-side cross section is well established \cite{Kho07r}
as the Airy oscillation of the nuclear rainbow pattern. The coupled channel (CC)
description of the purely elastic \oc scattering, {\it without} coupling to the $\alpha$ 
transfer channels, is shown in Fig.~\ref{fig:ES} where the cross section of the far-side 
scattering (\ref{NFdec}) has been obtained with 2 different absorption strengths 
of the complex OP from Table~\ref{t:OP}.
\begin{figure}[bhtp!]\vspace*{-1.5cm}\hspace*{0cm}
	\includegraphics[width=0.8\textwidth]{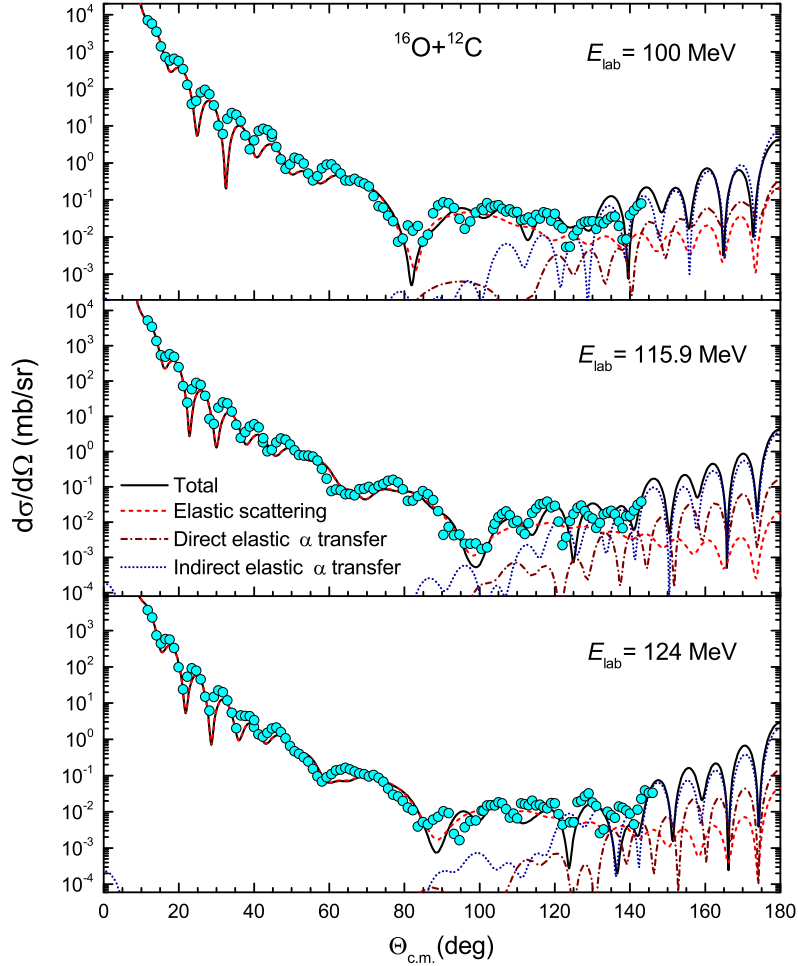}\vspace*{-1.5cm}
	\caption{4-channel CRC description (solid lines) of the elastic \oc data measured 
	at $E_\text{lab}=100, 115.9$, and 124 MeV \cite{Nic00} including the direct 
	and indirect elastic $\alpha$ transfer channels. The same complex OP (see 
	Table~\ref{t:OP}) was used in the CRC calculation of the purely elastic scattering 
	(dashed lines), direct $\alpha$ transfer (dash-dotted lines), and indirect $\alpha$ 
	transfer via the $2^+_1$ excited state of $^{12}$C (dotted lines).} 
 \label{fig:ET}
\end{figure}
One can see in Fig.~\ref{fig:ES} that the diffractive Fraunhofer oscillations 
at forward angles are followed immediately by a broad Airy oscillation of the 
elastic scattering cross section that is dominated by the far-side scattering. 
The deep minima of the elastic \oc scattering cross section observed in these data
were established \cite{Kho16,Phuc18} as the third Airy minimum (A3) at 
$\theta_\text{c.m.}\approx82^\circ$ in the energy of 100 MeV, and the second Airy minimum 
(A2) at $\theta_\text{c.m.}\approx98^\circ$ and $88^\circ$ in the energies of 115.9 
and 124 MeV, respectively. 
The renormalization factor $N_R$ of the real folded OP and WS parameters of the 
imaginary OP in Table \ref{t:OP} were fine tuned to reproduce both the Fraunhofer 
diffraction at forward angles and the broad Airy oscillation at medium and large 
angles, up to $\theta_\text{c.m.}\approx 100^\circ$. Since the coupling scheme considered in this work (Fig.~\ref{fig:Coup}) is different from those in Ref.~\cite{Phuc18}, a reproduction of experimental data requires different amounts of effective contributions from channels not explicitly included in the model space. Consequently, the fitted OP parameters from the two works have slightly different values to account for these scheme-specific contributions. 

Figure \ref{fig:ES} also illustrates that while the elastic data at forward angles are well described as the purely elastic scattering, the quick oscillation of the measured elastic scattering cross section at medium and large angles cannot be reproduced by the CC calculation in this work or the optical model (OM) description in Ref.~\cite{Phuc18}. Moreover, the leading-order Airy minima (A1 and A2) predicted by the OM analysis \cite{Kho16} at $\theta_\text{c.m.}>100^\circ$ are strongly deteriorated due to the contribution of the elastic $\alpha$ transfer in this angular range \cite{Phuc21}. 

The results of the 4-channel CRC calculation of the elastic \oc scattering with the
$\alpha$ transfer channels being included are shown in Fig.~\ref{fig:ET}, and one can see 
clearly the contribution of the elastic $\alpha$ transfer to the total elastic 
cross section at backward angles. The contribution of the indirect (two-step) elastic $\alpha$ transfer via 
the $2^+_1$ state of $^{12}$C is much stronger than the direct elastic $\alpha$ 
transfer one and dominates the elastic cross 
section at the most backward angles. This is due to a large spectroscopic factor 
$S_\alpha$ of the $\alpha + {\rm ^{12}C}^*_{2^+_1}$ configuration of the $^{16}$O 
nucleus discussed above. We reiterate that our CRC calculation does not treat the 
$S_\alpha$ factors as free parameters but adopts the $S_\alpha$ values predicted 
by the large-scale shell model calculation \cite{Vol15,Vol17}. The CRC results 
also show that the elastic $\alpha$ transfer occurring in the elastic \oc cross 
section at backward angles is well separated from the purely elastic scattering 
at the considered energies. The situation is different at lower energies \cite{Fer19} 
where the cross section of the purely elastic scattering is slightly larger and not as clearly distinct from that of the elastic $\alpha$ transfer at backward angles.  
\begin{figure}[bhtp!]\vspace*{-1.0cm}\hspace*{0cm}
	\includegraphics[width=0.8\textwidth]{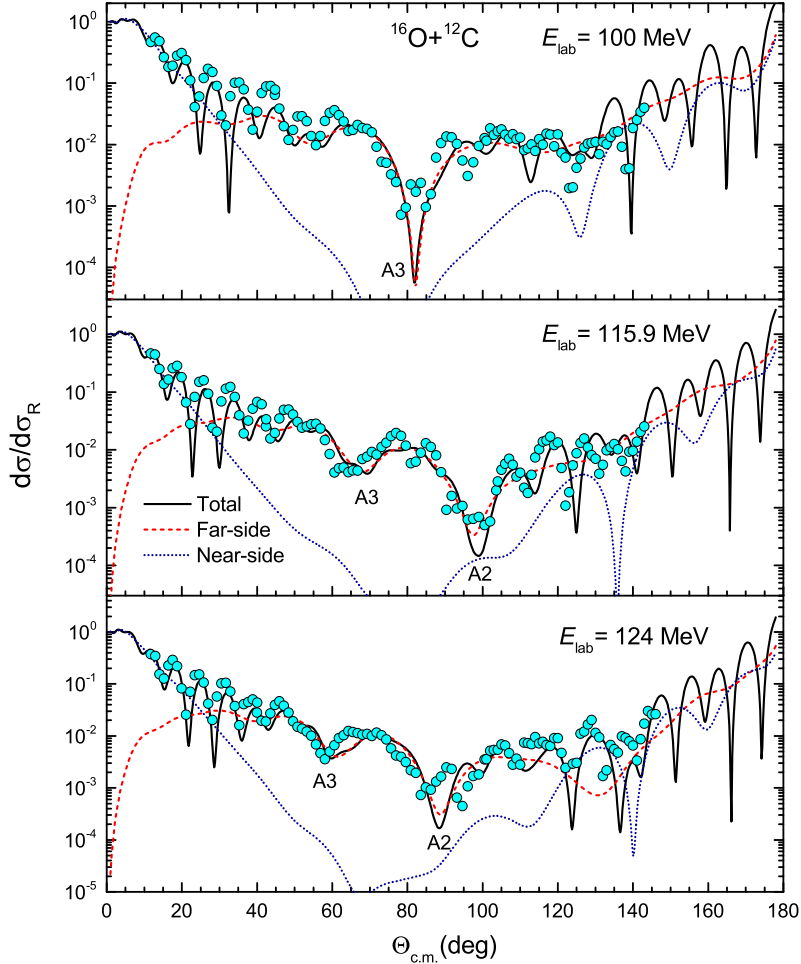}\vspace*{-1.5cm}
	\caption{Near-far decomposition  (\ref{NFdec}) of the CRC elastic \oc cross section 
	at $E_\text{lab}=100, 115.9$, and 124 MeV (solid lines) into the near-side (dotted 
	lines) and far-side (dashed lines) scattering cross sections using the Fuller's 
	method \cite{Ful75}, in comparison with the measured data \cite{Nic00}.} 
 \label{fig:NF-ET}
\end{figure}

The higher the incident energy the more elastic $\alpha$ transfer is separated from
the elastic scattering at medium and large angles, and this effect is more pronounced 
\cite{Phuc18} in the elastic \oc data at $E_\text{lab}=300$ MeV. Within the semiclassical 
interpretation of the elastic HI scattering \cite{Sat83}, more nonelastic channels are 
open with the increasing energy, and the absorption becomes, therefore, stronger and reduces 
the elastic flux at small partial waves, leading to a rapid decrease of the elastic 
scattering cross section at medium and large angles. A similar effect can be seen 
also in the cross section of the elastic $\alpha$ transfer. Because the elastic transfer 
amplitude at $\pi-\theta$ is added (\ref{eq:fsig}) to the elastic scattering amplitude 
at $\theta$, the elastic $\alpha$ transfer occurring mainly at the surface of the two 
colliding nuclei (or at forward angles) has the cross section largest at backward 
angles. We have further performed the near-far decomposition  (\ref{NFdec}) of the 
total CRC elastic \oc amplitude at the considered energies and the results are shown 
in Fig.~\ref{fig:NF-ET}. Because of the surface character of the $\alpha$ transfer,
the near-side cross section is naturally enhanced at large angles when the CRC 
coupling to the $\alpha$ transfer channels is included. It is interesting 
that a stronger near-side cross section given by the elastic $\alpha$ transfer 
amplitude at angles $\pi-\theta$ turns out to have a broad Airy-like oscillation 
from large to medium angles, which is likely associated with the (far-side)  
refracted $\alpha$ transfer waves evaluated at angles $\theta$. In conclusion, 
the observed oscillating distortion of the smooth Airy pattern of the nuclear 
rainbow in the elastic \oc scattering \cite{Phuc21} is due mainly to a stronger 
interference of the near-side and far-side scattering waves at medium and large 
angles caused by the (direct and indirect) elastic $\alpha$ transfer. 

\begin{figure}[bhtp!]\vspace*{-1.5cm}\hspace*{0cm}
	\includegraphics[width=0.8\textwidth]{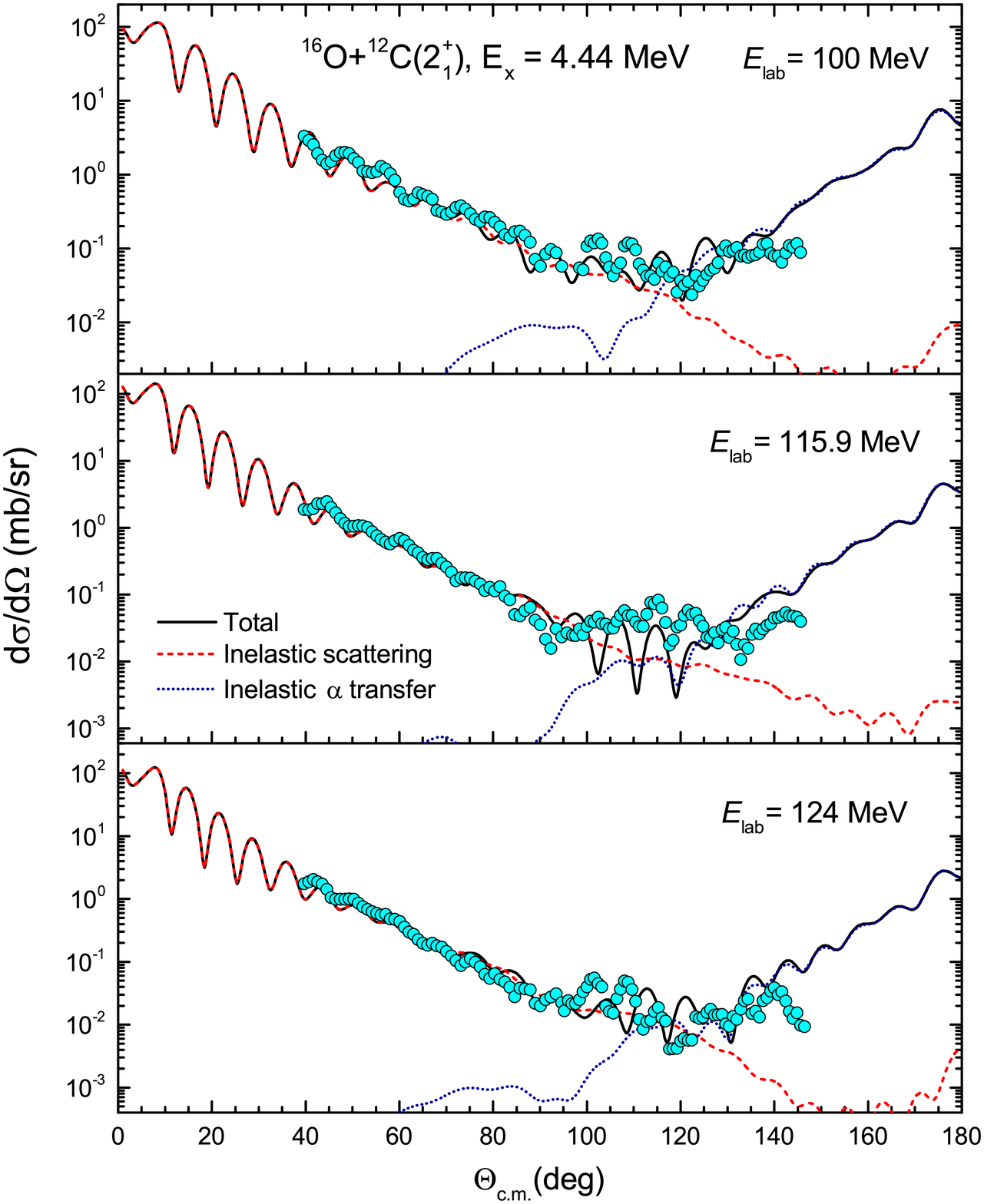}\vspace*{-1.5cm}
\caption{CRC description (solid lines) of the inelastic \oc scattering to the $2^+_1$ 
state of $^{12}$C in comparison with the data measured at $E_\text{lab}=100, 115.9$, 
and 124 MeV by Szilner {\it et al.} \cite{Szi06,Szilner}. The purely inelastic 
scattering (dashed lines) and inelastic $\alpha$ transfer (dotted lines) cross 
sections were obtained with the same OP as that used in the CRC calculation 
of the elastic \oc scattering (Table~\ref{t:OP}).} \label{fig:IET}
\end{figure}
The CRC results obtained within the 4-channel coupling scheme shown in Fig.~\ref{fig:Coup} 
describe well the data measured at the considered energies for the inelastic \oc 
scattering to the $2^+_1$ state of $^{12}$C by Szilner {\it et al.} \cite{Szi06,Szilner}
(see Fig.~\ref{fig:IET}). The purely inelastic scattering cross section at forward 
angles has the well established diffractive pattern that can be properly reproduced 
by the CC calculation {\it without} coupling to the $\alpha$ transfer channels 
(dashed lines in Fig.~\ref{fig:IET}). 
At variance with the elastic \oc scattering data where a pronounced rainbow pattern 
could be revealed at the three considered energies, a similar broad Airy oscillation
cannot be seen in the inelastic \oc scattering data for the $2^+_1$ state of $^{12}$C. 
Such a disappearance of the rainbow pattern in the inelastic \oc scattering has been
explained recently \cite{Phuc21-2} by extending the Fuller's decomposition method 
(\ref{NFdec}) for the near-far decomposition of the inelastic scattering amplitude. 
It was shown that a destructive interference of the coupled partial waves of different 
multipoles can suppress the oscillating Airy pattern in the inelastic scattering 
cross section when the nuclear excitation has nonzero spin \cite{Phuc21-2}. Such an 
effect is analogous to the opacity of an optical prism caused by the refraction 
of light rays having different wave lengths.   

Similar to the elastic \oc scattering considered above, the rise of the inelastic 
scatetring cross section at large angles is overwhelmingly dominated by the inelastic 
$\alpha$ transfer in the \oc system. A consistently good CRC description of both the 
elastic and inelastic \oc scattering data measured at $E_\text{lab}=100, 115.9$, and 
124 MeV \cite{Szi06,Szilner} has been obtained for the first time in the present 
work (see Figs.~\ref{fig:ET} and \ref{fig:IET}) with the large-angle data dominated 
by the elastic and inelastic $\alpha$ transfer. Like the elastic scattering case, the 
contribution of the inelastic $\alpha$ transfer to the inelastic scattering cross 
section becomes stronger with the increasing energies. At the considered energies, 
the inelastic $\alpha$ transfer contribution begins to dominate the inelastic scattering
cross section at $\theta_\text{c.m.}>120^\circ$ resulting in a distinctive V-shape
that cannot be properly described without coupling to the $\alpha$ transfer channels. 
Thus, our CRC results suggest that the elastic and inelastic \oc scattering data 
taken at higher incident energies of $E_\text{lab}> 100$ MeV are a suitable probe 
of the $\alpha$ cluster structure of $^{16}$O and $^{12}$C that can be revealed in 
the elastic and inelastic $\alpha$ transfer processes. 

\section{\Mr{Summary}}
The impact by the elastic and inelastic $\alpha$ transfer in the \oc system on the 
elastic scattering and inelastic scattering to the $2^+_1$ state of $^{12}$C 
has been thoroughly studied in a consistent 4-channel CRC analysis of the elastic
and inelastic \oc scattering data measured at the incident energies of $E_\text{lab}=100, 
115.9$, and 124 MeV. Our CRC calculations take explicitly into account the coupling 
between the elastic scattering, inelastic scattering, and $\alpha$ transfer channels, 
using the diagonal and transition \oc and \cc potentials obtained in the DFM 
calculation with the essential structure inputs like $S_\alpha$, overlap function, 
and nuclear transition densities taken from the reliable structure models. 

With the indirect $\alpha$ transfer via the $2^+_1$ excitation of the $^{12}$C core 
properly taken into account in the 4-channel CRC calculation of the elastic \oc 
scattering, a consistently good description of both the elastic and inelastic \oc 
data at the considered energies has been obtained for the first time. The contribution
from the elastic and inelastic $\alpha$ transfer is found to be dominant in the 
elastic and inelastic scattering cross sections at backward angles. The $\alpha$ transfer 
cross section at the considered energies is well separated from the purely scattering
cross section, at variance with that observed in the \oc data at the lower energy 
$E_\text{lab}\approx 80$ MeV \cite{Fer19}. Therefore, the results of the present CRC 
analysis suggest that the extensive elastic and inelastic \oc scattering data 
at refractive energies of $E_\text{lab}> 100$ MeV are more appropriate for the 
studies of elastic and inelastic $\alpha$ transfer.

\section*{\Mr{Acknowledgments}}
The present research was supported, in part, by the National Foundation for 
Scientific and Technological Development (NAFOSTED Project No. 103.04-2017.317). 
We thank Suzana Szilner for her helpful communication on the inelastic \oc
scattering data, and Alexander Volya for providing us with the $\alpha$ 
spectroscopic factors obtained from the large-scale shell-model calculation. 
The permission by Masayasu Kamimura to use the RGM nuclear transition 
densities in the present DFM calculation is also strongly appreciated.


\bibliographystyle{cip-sty-2019}
\bibliography{CIP-Submission}

\end{document}